\documentstyle[12pt,aps,tighten,epsf]{revtex}
\newcommand{\Ams}{{\protect\the\textfont2
   A\kern-.1667em\lower.5ex\hbox{M}\kern-.125emS}}
\begin{document}
\title{
Integrals of periodic motion
for classical equations of relativistic string with masses at ends}
\author{B.M.~Barbashov\address{Bogoliubov Laboratory of Theoretical Physics \\ JINR, 141980, Dubna,
Russia}}
\maketitle
\begin{abstract}
Boundary equations for the relativistic string with masses at ends
are formulated in terms of geometrical invariants of world trajectories of
masses at the string ends. In the three--dimensional Minkowski space $E^1_2$,
there are two invariants of that sort, the curvature $K$ and torsion $\kappa$.
For these equations of motion with periodic
$\kappa_i(\tau+n l)=\kappa(\tau)$, constants of motion are obtained.
\end{abstract}
\section{Equations of motion and boundary conditions}
Classical equations of motion and boundary conditions for a system of two
point masses connected by the relativistic string follow from the action
function for that system [1,2]
\begin{equation}
S=-\gamma \int d \tau \int d \sigma \sqrt{{\dot x\acute x}^{2}-{\dot x}^{2}
{\acute x}^{2}}
-\sum_{i=1}^2 m_i\int d \tau
\sqrt{{\left(
\frac{d x^{\mu}({\tau}_i,{\sigma}_i(\tau))}
                                           { d\tau}\right)}^{2}}
\label{a1}
\end{equation}
Here the first term is the action of a massless relativistic string; $\gamma$
is the parameter of tension of the string; $m_i$ are masses of particles at
the string ends; $x^{\mu}(\tau, \sigma)$ are coordinates of the string points
in a $D$--dimensional Minkowski space with metric $(1, -1, -1, ...)$;
derivatives are denoted by
\begin{eqnarray*}
{\dot x}^{\mu}=\partial x^{\mu}(\tau,\sigma)/\partial \tau, \qquad
{\acute x}^{\mu}=\partial x^{\mu}(\tau,\sigma)/\partial \sigma
\\
\frac{d x^{\mu}(\tau,\sigma_i(\tau))}{d \tau}=
\dot x^{\mu}(\tau,\sigma_i(\tau))+\acute x^{\mu}(\tau,\sigma_i(\tau))
\dot \sigma_i(\tau),
\end{eqnarray*}
where the string endpoints with masses in the plane of parameters $\tau$ and
$\sigma$ are described by functions $\sigma_i(\tau)$.

As in the case of a massless string, $m_i = 0$, the action (\ref{a1}) is
invariant with respect to a nondegenerate change of parameters
$\tilde \tau=\tilde \tau(\tau,\sigma)$ and
$\tilde \sigma=\tilde \sigma(\tau,\sigma)$, which allows us to take the
conformally flat metric on the string surface by imposing the conditions
of orthonormal gauge
\begin{equation}
\dot x^2+\acute x^2=0, \qquad \dot x \acute x=0
\label{a2}
\end{equation}

The action (\ref{a1})
results in the linear equations of motion for the string coordinates [1,2]
\begin{equation}
\ddot x^{\mu}(\tau,\sigma)-x''^{\mu}(\tau,\sigma)=0
\label{a3}
\end{equation}
and the boundary conditions for the ends with masses
$$
m_i \frac{d}{d \tau}
\left[ \frac{\dot x^{\mu}+
\dot \sigma_i
 \acute x^{\mu}}
{\sqrt{\dot x^2(1-{\sigma_i}^2)}}\right]
=
(-1)^{i+1}\gamma
\left[ \dot x^{\mu}+
\dot \sigma\acute x^{\mu}\right],
\label{a4}
$$
\begin{equation}
i=1,2;~x^{\mu}=x^{\mu}(\tau,\sigma_i(\tau));~\sigma_i=\sigma_i(\tau)
\end{equation}

The general solution to equations of motion (\ref{a3}) is the
vector--function
\begin{equation}
x^{\mu}(\tau,\sigma)=1/2\left[\psi^{\mu}_+(\tau+\sigma)+
\psi^{\mu}_-(\tau-\sigma)\right],\label{a5}
\end{equation}

Inserting it into the gauge conditions (\ref{a2}) we obtain the equations
$\acute \psi^2_+(\tau+\sigma)=0,~\acute \psi^2_-(\tau-\sigma)=0,$
where $\acute \psi^{\mu}_{\pm}(\tau\pm\sigma)$ are derivatives with respect to
the arguments.
According to (\ref{a5}) the vectors
$\acute \psi^{\mu}_+(\tau+\sigma) $
and
$\acute \psi^{\mu}_-(\tau-\sigma) $
should be isotropic.  For further consideration, it is convenient to represent
them as expansions over a constant basis in the $D$--dimensional Minkowski
space $E^1_{D_1}$ consisting of two isotropic vectors $a^{\mu}$ and
$c^{\mu}\,\, (a^{\mu}a_{\mu}=0,c^{\mu}c_{\mu}=0,a^{\mu}c_{\mu}=1)$
and $D - 2$ orthonormal space--like vectors $b^{\mu}_k\,\,(r=1,2,3...D-2),
b^{\mu}_k b_{l{\mu}}=-\delta_{kl}$
orthogonal to vectors $a^{\mu}$ and $c^{\mu}\,\,(a^{\mu}b_{k{\mu}}=0,
c^{\mu}b_{k{\mu}}=0)$~[2]. As
a result, we obtain the expansion of $\acute \psi^{\mu}_{\pm}$ over this
basis
\begin{eqnarray}
\acute \psi^{\mu}_+=\frac{A_+}
{\sqrt{\sum_{k=1}^{D-2} \dot f^2_k}}
\left[ a^{\mu}+\sum\limits_{k=1}^{D-2}b^{\mu}_k f_k+
\frac{1}{2}c^{\mu}\sum\limits_{k=1}^{D-2} f^2_k\right],\nonumber\\
\acute \psi^{\mu}_-=\frac{A_-}
{\sqrt{\sum_{k=1}^{D-2} \dot g^2_k}}
\left[ a^{\mu}+\sum\limits_{k=1}^{D-2}b^{\mu}_k g_k+
\frac{1}{2}c^{\mu}\sum\limits_{k=1}^{D-2} g^2_k\right],
\label{a7}
\end{eqnarray}
where $\acute\psi^{\mu}_{\pm}=\acute\psi^{\mu}_{\pm}(\tau\pm\sigma),
A_{\pm}=A_{\pm}(\tau\pm\sigma), f=f(\tau+\sigma),~g=g(\tau+\sigma)$.
It can easily be verified that $\acute \psi^2_{\pm}=0$, and
\begin{eqnarray}
\left(\psi''^{\mu}_{\pm}
\psi''_{\pm \mu}\right)=\psi''^2_{\pm}(\tau\pm\sigma)=
-A^2_{\pm}(\tau\pm \sigma),\nonumber
\end{eqnarray}
 where
$A^2_{\pm}(\tau\pm\sigma)$ are two arbitrary functions, like the functions
$f_k$ and $g_k$.  The condition of orthogonal gauge (\ref{a2}) does not
determine the functions $A_{\pm}$, and consequently, there is a possibility
to
fix them by imposing further gauge conditions since expressions (\ref{a7})
are
invariant under conformal transformations of the parameters $\tilde \tau \pm
\tilde \sigma=V_{\pm} (\tau \pm \sigma)$.  We fix them by imposing two more
gauge conditions
\begin{equation}
[\ddot x^{\mu}(\tau,\sigma)\pm \dot x'^{\mu}(\tau,\sigma)]^2=-A^2=
\mbox{const},
\label{a8}
\end{equation}
which in terms of the vector--functions
$\acute \psi^{\mu}_{\pm}$ mean
that the space--like vectors $\psi''^{\mu}_{\pm}(\tau\pm\sigma)$ are modulo
constant,
\begin{equation}
\psi''^2_{\pm}(\tau+\sigma)=-A^2_{\pm}(\tau\pm\sigma)=-A^2,
\end{equation}

In this way,
we have fixed the functions $A_{\pm}(\tau-\sigma)$ now equal to the constant
$A$.  At the same time, this condition fixes the values of functions
$\sigma_i(\tau)$ (see ref.[2] where it
is shown that $\sigma_i(\tau)=\sigma_i=const $), therefore, we choose
$\sigma_1(\tau)=0$ and $\sigma_2(\tau)=l $.

Further, we will consider the dynamics of a string with masses at the ends
on the plane $(x, y)$, i.e. in the Minkowski space with $D = 3$. In this
case,
the expansion (\ref{a7}) contains only one space--like vector $b^{\mu}$,
and the
expression (\ref{a7}) takes the form
\begin{eqnarray}
\acute \psi^{\mu}_+(\tau+\sigma)&=&\frac{A}{\acute f}
[a^{\mu}+b^{\mu}f+1/2 c^{\mu}f^2] \nonumber\\
\acute \psi^{\mu}_-(\tau-\sigma)&=&\frac{A}{\acute g}
[a^{\mu}+b^{\mu}f+1/2 c^{\mu}g^2],
\label{a9}
\end{eqnarray}
where $\dot f=\dot f(\tau+\sigma), \dot g=\dot g(\tau-\sigma)$ are
derivatives with respect to
arguments.

Boundary equations (1.4), when $\sigma_i(\tau) =$~const., $\dot x^{\mu}(\tau,
\sigma_i)$ and $\acute x^{\mu}(\tau,\sigma_i)$.
from (\ref{a5}) are substituted into them, and the representation (\ref{a9})
is taken into account, transform into two nonlinear equations for the
functions $f$ and $g$~[2]
\begin{eqnarray}
\frac{d}{d\tau}\ln \left[\frac{\acute g}{\acute f}\right]
+2 \frac{\acute f+\acute g}{f-g}&=&
\frac{\gamma}{m_1}\left|A\right|\frac{\left |f-g\right |}
{\sqrt{\acute f \acute g}}, \nonumber \\
\frac{d}{d\tau}\ln \left[\frac{\acute g_l}{\acute f_l}\right]
+2 \frac{\acute f_l+\acute g_l}{f_l-g_l}&
=&-\frac{\gamma}{m_2}\left|A\right|\frac{\left |f_l-g_l\right |}
{\sqrt{\acute f_l \acute g_l}},
\label{b1}
\end{eqnarray}
where $f=f(\tau),g=g(\tau),~f_l=f(\tau+l),~g_l=g(\tau-l)$. The nonzero
components of the metric tensor of the string surface
$\dot x^2(\tau,\sigma)=-\acute x^2(\tau,\sigma)$
are expressed via $f$ and $g$ as follows [2]
\begin{equation}
\dot x^2(\tau,\sigma)=A^2 \frac{[f(\tau+\sigma)-g(\tau-\sigma]^2}
{4 \acute f(\tau+\sigma)\acute g(\tau-\sigma)}.
\label{b2}
\end{equation}

From (\ref{b2}) we obtain the boundary values for the component of metric
tensor $\dot x^2(\tau,\sigma_i)$, where $\sigma_i=0,l$,
\begin{equation}
\dot x^2(\tau,0)=A^2 \frac{[f-g]^2}{4 \acute f\acute g},\;\;
\dot x^2(\tau,l)=A^2 \frac{[f_l-g_l]^2}{4 \acute f_l\acute g_l}.
\label{b3}
\end{equation}
By using these formulas together with eq.~(\ref{b1}) the functions $f(\tau)$
and $g(\tau)$ can be expressed~[2] in terms of the curvatures
$K_i$ and $\dot x^2(\tau,\sigma_i)$ as follows
\begin{eqnarray}
D\left[f\right]&=&I(\tau,\sigma_1)+Q(\tau,\sigma_1)
-2 K_1 A\frac{d}{d \tau}\sqrt{\dot x^2(0)} \nonumber \\
&=&I(\tau-l,\sigma_2)+Q(\tau-l,\sigma_2)
+2 K_2 A\frac{d}{d \tau}\sqrt{\dot x^2_{-l}(l)},
\label{b20} \\
D\left[g\right]&=&I(\tau,\sigma_1)+Q(\tau,\sigma_1)
+2 K_1 A\frac{d}{d \tau}
\sqrt{\dot x^2(0)}~~\nonumber\\
&=&I(\tau+l,\sigma_1)+Q(\tau+l,\sigma_2)
-2 K_2 A\frac{d}{d \tau}
\sqrt{\dot x^2_{+L}(l)}.
\label{b21}
\end{eqnarray}
Here
\begin{eqnarray}
I(\tau,\sigma_i)&=&D\left[A\int_{}^{\tau}\frac{d\eta}{\sqrt{\dot x^2(\eta,
\sigma_i)}}\right],\;\sigma_1=0,\sigma_2=l, \nonumber
\\
Q(\tau,\sigma_i)&=&\frac{A K_i}{2}\left(\frac{A}{K_i\dot x^2(\tau,\sigma_i)}
-\frac{K_i\dot x^2(\tau,\sigma_i)}{A}\right)
\end{eqnarray}
and $ D(f(\tau))$ stands for the Schwarz derivative:
$$D(f)=\frac{f'''}{f'}-\frac{3}{2}
\left(\frac{f''}{f'}\right)^2
=-2\sqrt{f'}\frac{d^2}{d\tau^2}\left(\frac{1}{\sqrt{f'}}\right).$$
The
second equalities in (\ref{b20}) and (\ref{b21}) represent just the
connection
between $\dot x^2(\tau,0)$ and $\dot x^2(\tau,l)$.

Further, from (\ref{b20}) and (\ref{b21}) it follows that the difference of
the Schwartz derivatives of the functions $f(\tau)$ and $g(\tau)$ is given by
\begin{eqnarray}
D[f]-D[g]&=&-4 A K_1 \frac{d}{d \tau}\sqrt{\dot x^2(0)},
\label{b22}\\
D[f_l]-D[g_l]&=&4 A K_2 \frac{d}{d \tau}\sqrt{\dot x^2(l)}.
\label{b23}
\end{eqnarray}
Eliminating $D[g(\tau)]$ from these equations by the change of $\tau$ to
$\tau +l$ in the  equation (\ref{b23}) and then eliminating $D[f(\tau)]$
by the change of
$\tau$ to $\tau - l$, we obtain the equations
\begin{eqnarray}
D[f_{2l}]-D[f]&=&4 A \frac{d}{d \tau}[K_1\sqrt{\dot x^2(0)}+
K_2\sqrt{\dot x^2_{+l}(l)}],\nonumber\\
D[g]-D[g_{2l}]&=&4 A \frac{d}{d \tau}[K_1\sqrt{\dot x^2(0)}+
K_2\sqrt{\dot x^2_{-l}(l)}],
\nonumber \\ \label{b24}
\end{eqnarray}
where $f_{2l}=f(\tau+2l), g_{2l}=f(\tau-2l), \dot x^2(0)=\dot x^(\tau,0),
\dot x^2_{-l}(l)=\dot x^2(\tau-l,l),\dot x^2_{+l}(l)=\dot x^2(\tau+l,l)$
The left--hand sides of (\ref{b24})
contain either the function $f$ or $g$ with shifted arguments, whereas the
right--hand sides depend on $\sqrt{\dot x^2(\tau,0)}$ and $\sqrt{\dot
x^2(\tau\pm l,l)}$.  These equations
give conserved quantities when the difference of Schwarz derivatives on the
 left--hand sides are zero under certain
conditions of periodicity to be considered in sect.~2.

\section{Constants of motion for boundary equations with periodic
torsions }

It is a remarkable fact that the system of boundary equations (\ref{b20}) and
(\ref{b21}) possesses conserved quantities when $\dot x^2(\tau,\sigma_i)$ are
periodic with a period multiple of $l$:  $\dot x^2(\tau,\sigma_i)=
\dot x^2(\tau+n l,\sigma_i), n=1,2,3,\ldots $; in this
case, the torsions of boundary curves will also be periodic,
$\kappa_i(\tau)=\kappa_i(\tau+n l)$.

The right--hand sides of the above
equations depend only on $\dot x^2(\tau,\sigma)$, consequently, their
left--hand sides should be periodic with the same period:
\begin{equation}
D[f_{n l}]=D[f],\quad
D[g_{n l}]=D[g],
\label{c1}
\end{equation}where $f_{nl}=f(\tau+nl),g_{nl}=g(\tau-nl)$
 In view of
the property of the Schwartz derivative we have
\begin{equation}
f_{n l}=\frac{a_1 f+b_1}{c_1 f+d_1}=T_1 f;~
g_{n l}=\frac{a_2 g+b_2}{c_2 g+d_2}=T_2 g.\label{c2}
\end{equation}
We
will prove that  these two linear-fractional
transformations are to be equal: $T_1=T_2$.
To this end, using (\ref{c2})
and (\ref{b3}), we write the condition of periodicity for $\dot
x^2(\tau,\sigma_i)$
\begin{eqnarray}
\dot x^2(0)&=&\dot x^2_{nl}(0)=\frac{A^2[T_1 f-T_2 g]^2}
{4 (T_1 f)'(T_2 g)'};\nonumber\\
\dot x^2(l)&=&\dot x^2_{+nl}(l)=\frac{A^2[T_1 f_l-T_2
g_{l}]^2} {4 (T_1 f_{l})'(T_2 g{l})'}.\label{c3}
\end{eqnarray}
Since the derivatives of the linear--fractional
function are given by the expressions
$$
\left(T_1 f\right)'=\frac{f'}{[c_1 f+d_1]^2},\quad
\left(T_2 g\right)'=\frac{g'}{[c_2 g+d_2]^2},
$$
 the denominators in (\ref{c3})
coincide, and the numerators obey the equality
$$
[f-g]=(a_1 f+b_1)(c_2 g+d_2)-(c_1 f+d_1)
(a_2 g+b_2)
$$ and the same equality
follows from the second eq. of (\ref{c3}) but with shifted arguments of
$f(\tau+l)$ and $g(\tau-l)$. These equalities, provided that
$a_i d_i-b_i c_i=
1$, hold valid under the condition
$a_1=a_2=a,~b_1=b_2=b,~ c_1=c_2=c,~d_1=d_1=d.$
Thus the periodicity condition (\ref{c3}) results in
that  $f$ and $g$ are transformed as follows
\begin{equation}
f_{n l}=T f,\quad g_{nl}=T g.
\label{c4}
\end{equation}

Now we can consider each of periods $l, 2l, ..., nl$ separately and
consequences  that follow from eqs. (\ref{b24}) in these cases.

For the period $l$, ${\dot x}^2(\tau+l,\sigma_i)={\dot x}^2(\tau,\sigma_i)$
from eq.~(\ref{c4}) it follows that $f(\tau+l)=Tf(\tau)$
and $g(\tau-l)=T^{-1}g(\tau)$, where $T^{-1}$ is the inverse
linear--fractional transformation, and
\begin{eqnarray*}
f_{2l}&=&T(T f)=\frac{(a^2+c b)f+b(a+d)}{c(a+d)f+d^2+c b},
 \\
g_{2l}&=&T^{-1}(T^{-1} g)=\frac{(d^2+c
b)g-b(a+d)}{-c(a+d)g+a^2+c b}
\end{eqnarray*}
 are also linear--fractional transformations with the determinant equal to
unity when $ad - bc = 1$. Then, the left--hand side of eqs. (\ref{b24}) are
zero because
$$D[f_{2l}]=D[f],\quad
D[g]=D[g_{2l}],
$$ we obtain the conserved quantity for the motion with
$
{\dot x}^2(\tau,\sigma_i)={\dot x}^2(\tau+l,\sigma_i)
$
\begin{equation}
K_1\sqrt{\dot x^2(0)}+K_2\sqrt{\dot x^2(l)}=h_1,
\label{c5}
\end{equation}
 where $h_1$ is the constant of integration.

Now let us consider the case with period $2l$: $\dot x^2(\tau+2l,\sigma_i)=
\dot x^2(\tau,\sigma_i)$.
According to (\ref{c4}), $f(\tau+2l)=T f(\tau)$ and $g(\tau-2l)
=T^{-1} g(\tau)$,
therefore,
$D[f(\tau+2l)]=D[f(\tau)]$
and
$D[g(\tau-2l)]=D[g(\tau)]$,
then the left-hand sides of eqs.  (\ref{b24}) again turn
out to be zero; upon integration we obtain
\begin{equation}
K_1\sqrt{\dot x^2(0)}
+K_2\sqrt{\dot x^2_{\pm l}(l)}= h_2.
\label{c6}
\end{equation}

From these examples it is not difficult to deduce the general expression for
a conserved quantity for period $nl$ that is different for even and odd $n$.

For even $n = 2r (r = 1, 2,...)$, it is necessary, upon adding $2ml$ to the
argument $\tau$ in eq. (\ref{b24}), to sum up the obtained expressions over
$m$
from zero to $r - 1$, which gives
$$
\sum\limits_{m=0}^{r-1}D[f_{2(1+m)l}]-
\sum\limits_{m=0}^{r-1}D[f_{2ml}]=
4A  \sum\limits_{m=0}^{r-1}\frac{d}{d \tau}
\left\{K_1 \sqrt{\dot x^2_{+2ml}(0)}+
K_2 \sqrt{x^2_{1+2ml}(l)}\right\}.
$$
The left--hand side of this equation equals zero since under the change
$1+m=m'$ in the first sum we have
$$
\sum_{m'=1}^{r}D[f_{2m'l}]-\sum_{m=0}^{r-1}D[f_{2ml}]=
D[f_{2rl}]-D[f]=0,
$$
and hence the constant quantity is
\begin{equation}
\sum_{m=0}^{r-1}\left\{K_1 \sqrt{\dot x^2_{+2ml}(0)}+
K_2 \sqrt{x^2_{l+2ml}(l)}\right\}=h_{2r}.
\label{c11}
\end{equation}
When $r = 1$, from (\ref{c11}) we obtain (\ref{c6}) with period $2l$.

For odd $n = 2r + 1 (r = 0, 1, 2, \ldots )$, it is necessary,
adding $ml$ to the
argument in (\ref{b24}), to sum up the equations over $m$ from zero to $2r$,
then
\begin{equation}
\sum_{m=0}^{2r}D[f_{2l+ml}]-\sum_{m=0}^{2r}D[f_{ml}]=
   4A \sum_{m=0}^{2r}\frac{d}{d \tau}\left\{K_1 \sqrt{\dot x^2_{+ml}(0)}+
K_2 \sqrt{x^2_{l+ml}(l)}\right\}.
\label{c12}
\end{equation}

Again, the left--hand side of the equation (3.12) is zero since setting
$2+m=m'$ in the first sum and considering that $(1+2k)l$ is a period, we get
$$\sum_{m'=2}^{2+2r}D[f_{m'l}]-\sum_{m=0}^{2r}D[f_{ml}]=
D[f_{(1+2r)l}]-D[f_{2(1+r)l}]-D[f]-D[f_l]=0,$$
Consequently, in this case the quantity
\begin{equation}
\sum_{m=0}^{2r}\left\{K_1 \sqrt{\dot x^2_{+ml}(0)}+
K_2 \sqrt{x^2_{+ml}(l)}\right\}=h_{2r+1}.
\label{c13}
\end{equation}
is constant.  In
(\ref{c13}) we considered that the last term in the sum of the second term
in (\ref{c12}) equals $\dot x^2(\tau+l+2rl,l)=\dot x^2(\tau,l)$. When $r = 0$
and $r = 1$, we obtain (\ref{c5}).

So, (\ref{c11}) and (\ref{c13}) are constants of motion of the boundary
equations of a relativistic string with masses at ends when masses are moving
along the curves with periodic torsion $\kappa_i(\tau+nl)=\kappa_i(\tau)$ and
constant curvature $K_i=\gamma/m_i$.
The work
is supported by the Russian Foundation for Fundamental Research (Grant 
No.~97--01--00745).

\end{document}